\documentclass[12pt]{article}
\usepackage{graphicx}

\thicklines

\newcommand{\EQ}{\begin{equation}}
\newcommand{\EN}{\end{equation}}
\newcommand{\bea}{\begin{eqnarray}}
\newcommand{\eea}{\end{eqnarray}}

\def\3i{\int\!\!\!\int\!\!\!\int}
\def\2i{\int\!\!\!\int}

\def\del{\Delta}
\def\ddel{{}^\bullet\! \Delta}
\def\deld{\Delta^{\hskip -.5mm \bullet}}

\def\ddeld{{}^{\bullet}\! \Delta^{\hskip -.5mm \bullet}}

\def\la{\langle}
\def\ra{\rangle}

\def\sqr#1#2{{\vcenter{\vbox{\hrule height.#2pt
     \hbox{\vrule width.#2pt height#1pt \kern#1pt
           \vrule width.#2pt}
       \hrule height.#2pt}}}}
\def\square{\mathchoice\sqr55\sqr55\sqr{2.1}3\sqr{1.5}3}
\begin{document}

\begin{flushright}
\begin{minipage}{0.25\textwidth} hep-th/0205182
\end{minipage}
\end{flushright}
\begin{center}
\bigskip\bigskip\bigskip
{\bf\Large{
Worldline formalism in a gravitational background}}
\vskip 1cm
\bigskip
Fiorenzo Bastianelli \footnote{E-mail: bastianelli@bo.infn.it} 
and
Andrea Zirotti \footnote{E-mail: zirotti@bo.infn.it} 
\\[.4cm]
{\em Dipartimento  di Fisica, Universit\`a di Bologna 
and  INFN, Sezione di Bologna\\ 
via Irnerio 46, I-40126 Bologna, Italy}
\end{center}
\baselineskip=18pt
\vskip 2.3cm

\centerline{\large{\bf Abstract}}
\vspace{.4cm}

We analyze the worldline formalism in the presence of a gravitational 
background. In the worldline formalism a path integral is used to quantize 
the worldline coordinates of the particles.
Contrary to the simpler cases of scalar and vector backgrounds, external 
gravity requires a precise definition of the ultraviolet regularization of 
the path integral. 
Taking into account the UV regularization, we describe the first quantized 
representation of the one-loop effective action for a scalar particle.  
We compute explicitly the contribution to the graviton tadpole and self-energy 
to test the validity of the method. The results obtained by usual field 
theoretical Feynman diagrams are reproduced in an efficient way.
Finally, we comment on the technical problems related to the factorization
of the zero mode from the path integral on the circle. 

\newpage

\section{Introduction}

The worldline path integral formulation of quantum field theory provides 
an alternative and efficient method for computing Feynman diagrams.  
This method has quite a long history \cite{Feynman50}. 
More recently, it has been developed further by viewing it as 
the particle limit of string theory \cite{BK}  and discussed directly as 
the first quantization of point particles \cite{Polyakov,Strassler}
(see \cite{Chris} for a review and a list of references).
In all these developments a difficult problem was the inclusion of gravity, 
even as a background.
In fact, gravity generically decouples in a naive particle limit of 
string theory, while in a direct worldline formulation the gravitational 
background leads to a path integral which necessarily requires a detailed 
discussion of ultraviolet regularizations.
Nevertheless, much progress has been made in 
\cite{Bern:1993wt} where string inspired rules were developed.

In this paper we address the use of the worldline formalism in the presence 
of background gravity starting directly from the first quantization
of point particles. 
We will describe how the interesting conceptual problems
posed by the gravitational background (see section 7.2 of ref. \cite{Chris})
can be dealt with.
We consider for simplicity the case of the one-loop 
effective action for a scalar particle in a gravitational background.  
We will show that by properly taking into account a precise ultraviolet
regularization scheme for the one-dimensional worldline path integral
allows one to include a gravitational background 
into the worldline formalism. 
 
The euclidean one-loop effective action $\Gamma[g]$ that we shall consider is 
the one obtained by quantizing a Klein-Gordon field $\phi$ coupled to gravity
\bea
S[\phi,g] 
= \int d^Dx \sqrt{g}\, {1\over 2} 
(g^{\mu\nu} \partial_\mu\phi \partial_\nu\phi +m^2\phi^2 +\xi R\phi^2 )
\label{uno}
\eea
and  formally reads 
($ e^{-\Gamma[g]} = \int {\cal D}\phi\ e^{-S[\phi,g]}$) 
\bea 
\Gamma[g] = {1\over 2} {\rm Tr} \log  (-\square +m^2 +\xi R)
\eea
with $\xi$ describing an additional non-minimal coupling.  

This effective action can also be obtained by considering the 
first quantization of a scalar point particle with 
coordinates $x^\mu$ and action \cite{Brink:1976sz}
\bea
S[e,x^\mu] = \int_{0}^{1} 
d\tau {1\over 2}
[e^{-1} g_{\mu\nu}(x) \dot x^\mu \dot x^\nu
+ e (m^2 + \xi R(x) )]
\eea
where $e$ is an auxiliary einbein and 
$ \dot x^\mu = \partial_\tau x^\mu $,
and requiring that the worldline is a closed loop (i.e. imposing periodic
boundary conditions for all fields).
By a standard gauge fixing procedure \cite{Polyakov}
one can eliminate the einbein by the gauge condition $e(\tau) = 2 T$ 
(the factor 2 is conventional and it is used here to obtain 
standard QFT formulas), thus
leaving an integration over the proper time parameter $T$
(the ghosts decouple and a factor ${1\over T}$ is due to the  
presence of an isometry on the circle)
\bea 
\Gamma[g] 
= -{1\over 2} \int_0^\infty {dT\over T } \int {\cal D}x\ e^{-S_{gf}[x^\mu]}
\label{effaction}
\eea
where 
\bea
S_{gf}[x^\mu] =
\label{sigma-model} 
\int_{0}^{1} 
d\tau 
\biggl ( 
{1\over 4 T} g_{\mu\nu}(x) \dot x^\mu \dot x^\nu +  T  (m^2 + \xi R(x) )
\biggr ) \ .
\label{gfaction}
\eea

Now one is left with a path integration over the coordinates $x^\mu $.
This is non-trivial as the nonlinear sigma model in (\ref{sigma-model})
needs a regularization.
Such nonlinear sigma models have been used previously to evaluate trace 
anomalies in 2, 4 and 6 dimensions
\cite{Bastianelli:1992be,Bastianelli:1993ct,Bastianelli:2000dw}, 
and in that context three different regularizations have been analyzed:
mode regularization  (MR) \cite{Bastianelli:1992be,Bastianelli:1998jm}, 
time slicing (TS) \cite{deBoer:1995hv}, and 
dimensional regularization (DR) \cite{Bastianelli:2000nm}.
The DR regularization was developed after the results 
of \cite{KC} which dealt with nonlinear sigma model 
in the infinite propagation time limit.
All these regularizations require different counterterms to produce 
the same physical results. 
The optimal choice for perturbative calculations is the 
DR scheme which requires a coordinate invariant counterterm 
\bea
\Delta S_{DR} = \int_{0}^{1} 
d\tau\, 2 T\, V_{DR}
\eea
to be added to (\ref{gfaction})
with $V_{DR}= -{1\over 8} R$. 
The MR and TS schemes on the contrary need each non-covariant counterterms, 
$V_{MR}$ and $V_{TS}$, that compensate the non-covariance of the 
regularization procedure.

A second technical issue concerns the ways one treats a constant zero
mode for the path integral on the circle. 
One option was already used in trace anomalies calculations. 
It consists in first considering loops with a fixed base-point $x^\mu_0$
in target space, and then integrating over the position of that base-point.
The coordinates $x^\mu(\tau)$ have Dirichlet boundary conditions
(DBC) $x^\mu(0) =x^\mu(1) =x^\mu_0$, so that the quantum fields 
$y^\mu(\tau)= x^\mu(\tau)-x^\mu_0$ describe fluctuations around
the background position $x^\mu_0$ which must vanish at $\tau =0,1$. 
These quantum fields have a kinetic term without zero modes
and the propagators can be derived immediately. 
This way of casting the path integral computation
delivers a covariant effective lagrangian density.
A second option, sometimes called ``string inspired'',
consists in directly separating out the constant zero mode 
$x^\mu_0 =\int_0^1 d\tau\, x^\mu (\tau)$ of the differential operator 
$\partial_\tau^2$ on the circle. 
The fields $y^\mu(\tau)= x^\mu(\tau)-x^\mu_0$ are now defined on the circle 
and thus satisfy periodic boundary conditions (PBC). 
The corresponding propagators are periodic and translationally invariant.
This set up is simpler than the first one since
in actual computations one can use translational invariance on the circle.
However, it has the disadvantage that it produces an   
effective lagrangian density with certain total derivative 
terms which are non-covariant. This non-covariance invalidates
any advantage of using Riemann normal coordinates
\cite{Schalm:1998ix}.
These two options are summarized in figures 1 and 2.
Other choices for treating the zero mode can be found in 
\cite{FT,Chris}.
The total derivative terms
of the ``string inspired'' method are present not only in
the gravitational case. They exist also for standard field theories 
in flat space, including gauge theories,
but in that case they are not bothersome, since 
they do not violate gauge invariance. In fact, they are 
even beneficial since their addition leads to a more compact
form of the effective action \cite{Fliegner:1997rk}.

\begin{figure}[ht]
\centering
\includegraphics{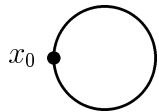}
\caption{\em Dirichlet boundary conditions at $x_0$ (DBC)}
\label{fig1}
\end{figure}

\begin{figure}[ht]
\centering
\includegraphics{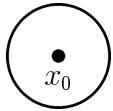}
\caption{\em Periodic boundary conditions without zero mode (PBC)}
\label{fig2}
\end{figure}

In the following we shall consider the one-loop effective action as the 
generator of 1PI graphs and we shall evaluate directly 
the terms obtained after functional differentiation.
In such a situation total derivative terms which may be present 
in the effective lagrangian density are harmless. Thus the simplest way to
set up the computations in the worldline scheme is to use the 
``string inspired'' method for separating out the zero mode together with 
worldline dimensional regularization. This way we will obtain the
correct contribution to the tadpole and self-energy of the graviton,
showing the correctness and 
efficiency of the worldline formalism in this context.

As already mentioned, previous results in the string inspired framework 
for describing theories coupled to gravity were presented in 
\cite{Bern:1993wt}, and more recently in \cite{Bern:1998ug} where 
perturbative formulations of certain gravitational theories have been 
systematically carried out by using relations between open and 
closed string amplitudes. 
In the present paper we only consider external gravity, but it would be
quite interesting to extend our results by learning how
to include other particles as well as gravity itself in the loops,
and possibly making contact with the 
rules developed in \cite{Bern:1993wt,Bern:1998ug}.

The paper is organized as follows. In section 2 we 
describe the  worldline formalism with background gravity.
In section 3 we present explicit computations of the one- and 
two-point functions, which give the scalar particle 
contributions to the cosmological constant and graviton self-energy,
respectively.
In section 4 we test the correctness of our results by 
first showing that the expected Ward identities are satisfied, 
and then by comparing with a standard Feynman graph calculation.
In section 5 we discuss the total derivative terms that one finds
in computing directly the effective lagrangian density 
with the PBC (``string inspired'') propagators.
Finally we present our conclusion in section 6 and put in an appendix
our conventions and some useful formulae. 


\section{The worldline formalism}

The worldline formalism for a scalar particle in a gravitational
background leads us to consider the following 
representation of the one-loop effective action
\bea 
\Gamma[g] 
= - {1\over 2} \int_0^\infty {dT\over T } \int {\cal D}x\ e^{-S_{gf}[x^\mu]}
\label{effaction2}
\eea
where 
\bea
S_{gf}[x^\mu] = \int_{0}^{1} d\tau 
\biggl ( {1\over 4 T} 
g_{\mu\nu}(x) \dot x^\mu \dot x^\nu +  T (m^2 + \xi R(x))
\biggr ) 
\label{gfaction2}
\eea
and with the fields $x^\mu(\tau)$ satisfying periodic boundary conditions
at $\tau=0,1$.
Because of derivative interactions present in this nonlinear sigma
model divergences may arise in the quantum-mechanical loop corrections. 
Thus one needs a regularization.
However one does not need infinite renormalization:
the covariant path integral measure
produces other infinities that cancel the original ones
\cite{ly}.
This measure is of the form
\bea
{\cal D} x = Dx \prod_{ 0\leq \tau < 1} \sqrt{\det g_{\mu\nu}(x(t))}   
\eea
where $Dx=\prod_\tau d^Dx(\tau)$ denotes
the standard translationally invariant measure. 
It can be represented more conveniently by introducing
bosonic $a^\mu$ and fermionic $b^\mu , c^\mu$ ghosts which satisfy the 
same periodic boundary conditions of the coordinates $x^\mu$ 
\cite{Bastianelli:1992be,Bastianelli:1993ct}  
\bea
{\cal D} x = Dx \prod_{ 0 \leq \tau < 1} \sqrt{\det g_{\mu\nu}(x(t))}   =
Dx \int { D} a { D} b { D} c \;
{\rm e}^{- S_{gh}[x,a,b,c]} 
\eea
where
\bea
S_{gh}[x,a,b,c]
= \int_{0}^{1} d\tau \; {1\over 4T}g_{\mu\nu}(x)(a^\mu a^\nu 
+ b^\mu c^\nu) \ . 
\eea
After having selected a regularization scheme one may explicitly check 
that all divergences cancel and only certain spurious finite terms are 
left over.
The latter are compensated by a finite counterterm $V_{CT}$ associated with 
the chosen regularization scheme.

Here we will adopt dimensional regularization
which was shown in \cite{Bastianelli:2000nm} 
to require the covariant counterterm
\bea
\Delta S_{DR}[x] = \int_{0}^{1} 
d\tau\, 2 T\, V_{DR}(x)
\eea
with $V_{DR}= -{1\over 8} R$. 
In dimensional regularization, as developed in \cite{Bastianelli:2000nm}, 
one may proceed as follows: {\it (i)} 
one extends the original compact
space $I=[0,1]$ by adding $d$ infinite dimensions, {\it (ii)} 
uses partial integration in the regulated $d+1$ dimensions
to cast in simpler forms the integrals arising in perturbation theory, 
{\it (iii)} 
computes those simpler forms by first removing the regularization 
(i.e. sending $d\to 0$) in case that no ambiguities are left over at 
$d=0$ \cite{Bastianelli:2000nm,Bastianelli:2000dw}.
This procedure frees one from the need of computing tricky integrals 
at arbitrary complex $d+1$ dimensions, as done instead in the usual QFT 
dimensional regularization. 
In all the cases analyzed so far this recipe has been enough to compute
the required integrals.

Collecting all terms, the formula for the effective action
in the worldline DR scheme is given by
\bea 
\Gamma[g] 
= - {1\over 2} \int_0^\infty {dT\over T } \int Dx Da Db Dc
\ e^{-S}
\label{effaction3}
\eea
with
\bea
S= \int_{0}^{1} d\tau 
\biggl ( {1\over 4 T} 
g_{\mu\nu} (\dot x^\mu \dot x^\nu + a^\mu a^\nu +b^\mu c^\nu)
+  T (m^2 + \bar \xi  R)
\biggr ) 
\label{gfaction3}
\eea
where 
$\bar \xi =\xi -{1\over 4} $ takes into account the DR counterterm.

This effective action can be used immediately to obtain (1PI) correlation 
functions by varying the metric $g_{\mu\nu}$ and then setting
$g_{\mu\nu}=\delta_{\mu\nu}$. Alternatively, one  
can obtain correlation functions directly in momentum space.
One considers the effective action as a power series in 
$h_{\mu\nu}= g_{\mu\nu}-\delta_{\mu\nu}$,
substitutes the $h^N$ term with plane waves of definite polarizations
\bea
h_{\mu\nu} (x)
= \sum_{i=1}^{N} \epsilon_{\mu\nu}^{(i)} 
e^{ip_i \cdot x} 
\label{quindici}
\eea
and then picks up the terms linear in each $\epsilon^{(i)}_{\mu\nu} $:
this gives the contribution to the $N$-graviton amplitude
in momentum space $\tilde \Gamma^{\epsilon_1,..,\epsilon_N}_{(p_1,..,p_N)}$
(see notation in eq. (\ref{amplitude})).

This way one is left with quantum mechanical correlation functions 
on the circle of the form
\bea
\biggl \la ( \dot x^{\mu_1}_1 \dot x^{\nu_1}_1 + 
a^{\mu_1}_1 a^{\nu_1}_1 + b^{\mu_1}_1 c^{\nu_1}_1) e^{ip_1 \cdot x_1 } 
\cdots
(\dot x^{\mu_N}_N \dot x^{\nu_N}_N + 
a^{\mu_N}_N a^{\nu_N}_N + b^{\mu_N}_N c^{\nu_N}_N) e^{ip_N \cdot x_N } 
\biggr \ra 
\label{sedici}
\eea
where the fields $x_1, a_1$ stand for $x(\tau_1), a(\tau_1) $ and so on
(this formula is exact for $\bar \xi=0$, the general case has additional 
contact terms due to vertices with multiple graviton legs arising from the 
expansion of the $\bar \xi R$ term). 

On the circle the free kinetic term of the coordinates $x^\mu$ has a 
zero mode. One option is to split 
\bea
x^\mu(\tau) =x^\mu_0 +y^\mu(\tau), \ \ \ \ \ \ \ 
y^\mu(\tau) =\sum_{n\neq 0} y_n^\mu e^{2 \pi i  n \tau}
\eea
where $x^\mu_0$ is the constant zero mode of the differential operator
$\partial_\tau^2$ and $y^\mu(\tau)$ are the quantum fluctuations.
After inclusion of the quantum fluctuations one must
integrate over all possible zero modes, i.e. all possible positions
of the particle loop in target space.
Thus the path integration is split as
\bea
 Dx= {1\over (4 \pi T)^{D \over 2}}\, d^Dx_0\,  Dy \ .
\label{mesplit}
\eea
The kinetic term for the quantum fields $y^\mu$ is now invertible and 
the corresponding path integral is normalized to unity
\bea
\int Dy\ e^{-\int_0^1 d\tau {1\over 4 T} \dot y^2} =  1 \ .
\eea
The propagators are translationally invariant and read
\bea
\la y^\mu (\tau) y^\nu(\sigma)\ra \!\! &=& \!\!
- 2 T\ \delta^{\mu\nu}\ 
\Delta(\tau-\sigma)
\nonumber\\
\la a^\mu(\tau) a^\nu(\sigma)\ra \!\! &=& \!\!
  2T\ \delta^{\mu\nu}\ \Delta_{gh} (\tau- \sigma) 
\nonumber\\
\la b^\mu(\tau) c^\nu (\sigma)\ra \!\! &=& \!\! -4T\ \delta^{\mu\nu}\ 
\Delta_{gh}(\tau-\sigma)
\label{propag}
\eea
where $\Delta$ and $\Delta_{gh}$ are given by
\bea
&& \Delta (\tau-\sigma) = -\sum_{n\neq0}
 {1 \over {4 \pi^2 n^2}} e^{2 \pi i  n (\tau -\sigma )}
={1\over 2} |\tau-\sigma| -
{1\over 2} (\tau-\sigma)^2 -{1 \over 12}
\nonumber\\
&& \Delta_{gh}(\tau-\sigma) = \sum_{n=-\infty}^{\infty}
e^{2 \pi i  n (\tau -\sigma )} = \delta(\tau-\sigma) 
\ .
\label{ghost-prop}
\eea
With these propagators one can compute the averages in (\ref{sedici}) 
using the Wick theorem.

Note that integration over the zero mode $d^Dx_0$ in (\ref{mesplit})
produces through the exponentials in (\ref{sedici}) a delta function 
\bea
(2 \pi)^D \delta^D (p_1 +\cdots + p_N) 
\eea
enforcing momentum conservation. 
With momentum conservation the constant part of the propagator
$\Delta$ drops out from (\ref{sedici}) and may be set to zero.
Thus instead of $\Delta $ one may use the effective propagator
\bea
\Delta_0 (\tau-\sigma) =
{1\over 2} |\tau-\sigma| - {1\over 2} (\tau-\sigma)^2 
\eea
which satisfies $\Delta_0 (0) =0$.

We will apply and test this set up to compute one- and two-point 
functions in the next section.

\section{One- and two-point functions}

We begin with the rather simple one-point function which gives the
scalar particle contribution to the cosmological constant in figure 3.
Taking from the effective action the term linear in  $h_{\mu\nu}$ 
and substituting for $h_{\mu\nu}$  the expression 
(\ref{quindici}) with just one plane wave produces
\bea
\tilde \Gamma^{\epsilon}_{(p)}
= {1\over 2} \int_{0}^{\infty} 
{dT\over T} e^{-m^2 T} {1\over (4 \pi T)^{D\over 2} } 
\int d^Dx_0 
{1\over 4 T }  \epsilon_{\mu\nu} \int_0^1 \! d\tau \,
\biggl \la ( \dot y^{\mu} \dot y^{\nu} + 
a^{\mu} a^{\nu} + b^{\mu} c^{\nu}) e^{ip \cdot (x_0 + y)} 
\biggr \ra \ .
\eea
\begin{figure}[ht]
\centering
\includegraphics{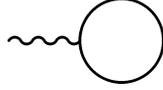}
\caption{\em Graviton tadpole}
\label{fig3}
\end{figure}

\noindent The integration over $d^Dx_0$ gives the momentum delta function
$(2 \pi)^D \delta^D(p) $, which for simplicity
we factorize together with the polarization tensor $\epsilon_{\mu\nu}$. 
Momentum conservation eliminates the exponential 
and the remaining Wick contractions leave us with\footnote{Here we consider 
$\Delta (\tau,\sigma) =\Delta (\tau-\sigma)$ 
as  function of two variables, and dots on the left/right 
denote derivatives with respect to $\tau$/$\sigma$.
Later on we will also denote by $\Delta|_\tau$ evaluation at 
coinciding points $\sigma =\tau$.}
\bea
\Gamma^{\mu\nu}_{(0)} = -{\delta^{\mu\nu} \over 4}  
\int_{0}^{\infty} 
{dT\over T} e^{-m^2 T} {1\over (4 \pi T)^{D\over 2} } 
\int_0^1 \! d\tau  \,
( \ddeld(\tau-\tau) + \Delta_{gh}(\tau-\tau)) \ .
\eea
The integrals of the propagators can be treated in DR,
but it is immediately clear from eq. (\ref{ghost-prop}) that 
only the term from the ghost zero mode 
contributes
\bea
\ddeld(\tau-\tau) + \Delta_{gh}(\tau-\tau) = 1 \ .
\label{onepoint}
\eea
Here we see explicitly the effect of the ghosts that eliminate 
potential divergences in quantum mechanics. 
Now the integral over the proper time leads directly to a gamma
function if the target space dimension $D$
is turned into a complex number 
(this is dimensional regularization in target space
and regulates the QFT ultraviolet divergences).
So we are left with the result
\bea
\Gamma^{\mu\nu}_{(0)} = -{\delta^{\mu\nu} \over 4}  
{(m^2)^{D\over 2}\over (4 \pi)^{D\over 2} } 
\Gamma\Big(-{D\over 2}\Big) \ .
\label{1ptf}
\eea
Note that the terms linear in $h_{\mu\nu}$ and coming from
the expansion of the scalar curvature in (\ref{gfaction3})
vanish at zero momentum. 
This tadpole diagram diverges at even dimensions $D$
and must be renormalized. Of course, one may keep $D$ fixed
and use instead a cut-off in the proper time 
as an alternative regularization.

We now discuss the more interesting two-point function.
Let's consider first the simpler case with $\bar \xi = 0$
(i.e. $\xi = {1\over 4}$). 
This is special since only vertices with one graviton are present,
see figure 4.
\begin{figure}[ht]
\centering
\includegraphics{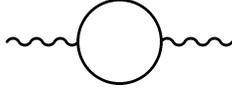}
\caption{\em Graviton self-energy}
\label{fig4}
\end{figure}
The term of the effective action 
quadratic in the metric fluctuations $h_{\mu\nu}$ is
\bea
\tilde \Gamma^{\epsilon_1,\epsilon_2}_{(p_1,p_2)}
= - {1\over 2} \int_{0}^{\infty} 
{dT\over T} e^{-m^2 T} {1\over (4 \pi T)^{D\over 2} } 
\int d^Dx_0\, 
\biggl \la {1\over  2 }   
\biggl (\int_0^1 d\tau {1\over 4 T}
 h_{\mu\nu}( \dot y^{\mu} \dot y^{\nu} + 
a^{\mu} a^{\nu} + b^{\mu} c^{\nu})
\biggr )^2 
\biggr \ra  \biggr|_{lin\ \epsilon_1, \epsilon_2}
\eea
where 
\bea
h_{\mu\nu} 
= \epsilon_{\mu\nu}^{(1)} e^{ip_1 \cdot x}
+ \epsilon_{\mu\nu}^{(2)} e^{ip_2 \cdot x} \ .
\eea
As before the zero mode integration gives a delta function for
momentum conservation, which we factorize again for notational simplicity. 
Then a straightforward application of the Wick theorem produces
\bea
\Gamma^{\epsilon_1\epsilon_2}_{(p,-p)} 
= - {1\over 8}  {1\over (4 \pi)^{D\over 2}}
\int_{0}^{\infty} 
{dT\over T^{1 + {D\over 2}}} e^{-m^2 T} ( 
 r_1 I_1 + r_2 I_2
- 2 T p^2 (r_3 I_3 - r_4 I_4)
+ 4 T^2 p^4 r_5 I_5 )
\eea
where $p=p_1=-p_2$ and
$r_i = \epsilon^{(1)}_{\mu\nu} 
R_i^{\mu\nu\alpha\beta} \epsilon^{(2)}_{\alpha \beta} $
with
\bea
R_1^{\mu\nu\alpha\beta} \!\! &=& \!
\delta^{\mu\nu}\delta^{\alpha\beta} 
\nonumber \\
R_2^{\mu\nu\alpha\beta} \!\! &=& \!
\delta^{\mu\alpha}\delta^{\nu\beta}+ \delta^{\mu\beta}\delta^{\nu\alpha} 
\nonumber \\
R_3^{\mu\nu\alpha\beta} \!\! &=& \!
{1\over p^2} \, (\delta^{\mu\alpha} p^\nu p^\beta +
\delta^{\nu\alpha} p^\mu p^\beta +
\delta^{\mu\beta} p^\nu p^\alpha +
\delta^{\nu\beta} p^\mu p^\alpha )
\nonumber \\
R_4^{\mu\nu\alpha\beta} \!\! &=& \!
 {1\over p^2} \, (
\delta^{\mu\nu} p^\alpha p^\beta
+\delta^{\alpha\beta} p^\mu p^\nu) 
\nonumber \\
R_5^{\mu\nu\alpha\beta} \!\! &=& \!
 {1\over p^4}\, p^\mu p^\nu p^\alpha p^\beta 
\eea
while the integrals coming from the quantum mechanical 
correlation functions are given by
\bea
I_1 \!\! &=& \!\!
 \int_0^1 \!\! d\tau \!\! \int_0^1 \!\! d\sigma\ 
(\ddeld + \Delta_{gh})|_\tau \, (\ddeld + \Delta_{gh})|_\sigma \
e^{-2 T p^2 \Delta_0} 
\nonumber \\[1mm]
I_2 \!\! &=& \!\!
 \int_0^1 \!\! d\tau \!\! \int_0^1 \!\! d\sigma\ 
(\ddeld{}^2  -\Delta_{gh}^2)\ 
e^{- 2 T p^2 \Delta_0}
\nonumber \\[1mm]
I_3 \!\! &=& \!\!
 \int_0^1 \!\! d\tau \!\! \int_0^1 \!\! d\sigma\ 
\ddel\, \ddeld \, \deld  \,
e^{- 2 T p^2 \Delta_0}
\nonumber \\[1mm]
I_4 \!\! &=& \!\!
 \int_0^1 \!\! d\tau \!\! \int_0^1 \!\! d\sigma\ 
(\ddeld + \Delta_{gh})|_\tau\, (\deld)^2 \,
e^{- 2 T p^2 \Delta_0}
\nonumber \\[1mm]
I_5 \!\! &=& \!\!
 \int_0^1 \!\! d\tau \!\! \int_0^1 \!\! d\sigma\ 
(\ddel)^2\, (\deld )^2 \,
e^{- 2 T p^2 \Delta_0} \ .
\label{integrals}
\eea
Translational invariance can be used at once by fixing $\sigma =0$.
Then one obtains the following results by 
using dimensional regularization when necessary 
\bea
I_1 \!\! &=& \!\!
\int_0^1  d\tau \ e^{- T p^2 (\tau-\tau^2)} \nonumber \\[1mm]
I_2 \!\! &=& \!\!
 {1\over 4} T p^2  -2 + I_1 \nonumber \\[1mm]
I_3 \!\! &=& \!\!
 {1\over 8} -{1\over 2 T p^2 }(1-I_1) \nonumber \\[1mm]
I_4 \!\! &=& \!\!
{1\over 2 T p^2 }(1-I_1) \nonumber \\[1mm]
I_5 \!\! &=& \!\!
 {1\over 8 T p^2 }-{3\over 4 T^2 p^4}(1-I_1) \ .
\label{values}
\eea
At this stage the proper time integral can be carried out
at complex $D$ and yields 
\bea
(4 \pi )^{D\over 2} \Gamma_{(p,-p)} \!\! &=& \!\!
-{1\over 8} \Gamma\Big(-{D\over 2}\Big)
\biggl [  
(P^2)^{{D\over 2}}
(R_1 + R_2 - R_3 - R_4 + 3 R_5 )
- (m^2)^{{D\over 2}} (2 R_2 - R_3 - R_4 + 3 R_5 ) \biggr ]  
\nonumber \\[2mm] 
\!\! &-& \!\!
{1\over 32} \Gamma\Big(1-{D\over 2}\Big) p^2 (m^2)^{{D\over 2}-1} 
( R_2  - R_3 + 2 R_5 ) 
\label{2ptf}
\eea
where in our shorthand notation we have factorized the polarization 
tensors $\epsilon^{(i)}_{\mu\nu}$, 
suppressed tensor indices, and used the expression
\bea
(P^2)^x = \int_0^1 d\tau\, (m^2 + p^2 (\tau -\tau^2))^x \ .
\eea 

The additional terms $\Delta \Gamma_{(p,-p)} $
present for the case $\bar \xi \neq 0 $ correspond to figure 5 and 
can be quickly derived by using the expansion of the scalar curvature
reported in eq. (\ref{r2}) of the appendix. They read
\bea
 (4 \pi )^{D\over 2} \Delta \Gamma_{(p,-p)} \!\! &=& \!\!
-{\bar\xi  \over 8} 
\Gamma\Big(1-{D\over 2}\Big)
p^2 \biggl [ (m^2)^{{D\over 2}-1} 
 (2 R_1  +  R_2 -  R_3 - 2 R_4  + 4  R_5 ) 
\nonumber \\[2mm] 
&& \hspace{-2cm}
-4 (P^2)^{{D\over 2}-1} ( R_1  -  R_4  +  R_5 ) 
\biggr]
-{\bar\xi^2 \over 2} 
\Gamma\Big(2-{D\over 2}\Big) 
p^4 (P^2)^{{D\over 2}-2} (R_1 - R_4  +R_5) \ . 
\label{add}
\eea
\begin{figure}[hb]
\centering
\includegraphics{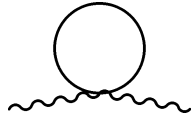}
\caption{\em Additional graph for graviton self-energy}
\label{fig5}
\end{figure}

\section{Ward identities and standard Feynman graphs}

Ward identities follow from general coordinate invariance and
can be used to test our previous results. 
General coordinate invariance implies conservation of the induced stress 
tensor 
\bea
&&\nabla_\mu^{(x)} {1\over \sqrt{g(x)}} 
{\delta \Gamma[g] \over \delta g_{\mu\nu}(x)} =0 \ .
\label{wwi}
\eea
We use the notation
\bea
{\delta^n \Gamma [g] \over \delta g_{\mu_1\nu_1}(x_1) . . 
\delta g_{\mu_n\nu_n}(x_n)} 
\bigg |_{g_{\mu\nu}=\delta_{\mu\nu}} \equiv
\Gamma^{\mu_1\nu_1,..,\mu_n\nu_n}_{(x_1,..,x_n)}
\label{notation}
\eea
and Fourier transform to momentum space by
\bea
\tilde \Gamma_{(p_1,.., p_n)} =
(2\pi)^D \delta(p_1+..+p_n)\Gamma_{(p_1,.., p_n)} =
\int dx_1 .. dx_n\, e^{ip_1x_1 +.. +ip_nx_n}\, \Gamma_{(x_1,.., x_n)} \ .
\label{amplitude}
\eea
Thus from (\ref{wwi}) we get the following Ward identity relating the 
one- and two-point functions
\bea
p_\mu \Gamma^{\mu\nu,\alpha\beta}_{(p,-p)}
+{1\over 2} p_\mu (\delta^{\nu\beta} \Gamma^{\mu\alpha}_{(0)} 
+ \delta^{\nu\alpha} \Gamma^{\mu\beta}_{(0)} )
-{1\over 2} p^\nu \Gamma^{\alpha\beta}_{(0)} =0 \ .
\eea
It is immediate to verify from eqs. (\ref{1ptf}), (\ref{2ptf}),
(\ref{add}) that this identity is satisfied for any value of $\xi$.

One may note that the two-point function can be written in a 
more compact form which makes it easier to check the Ward identity.
Defining the tensors
\bea
S_1 \!\! &=& \!\! 
R_1 -  R_4 + R_5 \cr
S_2 \!\! &=& \!\! 
R_2 -  R_3 + 2 R_5 
\eea
which satisfy
$p_\mu S_1^{\mu\nu\alpha\beta} = p_\mu S_2^{\mu\nu\alpha\beta} = 0$
allows one to write the full two-point function as
\bea
&& (4 \pi )^{D\over 2} \Gamma^{full}_{(p,-p)} =
-{1\over 8} \Gamma\Big(-{D\over 2}\Big) 
 \biggl [ (m^2)^{{D\over 2}} ( R_1  -R_2 ) 
+ ((P^2)^{{D\over 2}} -(m^2)^{{D\over 2}})(S_1  +S_2) 
\biggr] \nonumber \\[2mm] && 
- {1\over 32} \Gamma\Big(1-{D\over 2}\Big)\, 
p^2 \biggl [  
(m^2)^{{D\over 2}-1} S_2 \biggr ]
-{\bar\xi \over 8} 
\Gamma\Big(1-{D\over 2}\Big) \, 
p^2 \biggl [ (m^2)^{{D\over 2}-1} ( 2 S_1  + S_2 ) 
- 4 (P^2)^{{D\over 2}-1}  S_1  \biggr] 
\nonumber \\[2mm] && 
-{\bar\xi^2 \over 2} 
\Gamma\Big(2-{D\over 2}\Big) \, 
p^4  \biggl [  
(P^2)^{{D\over 2}-2} S_1 \biggr ] \ .
\label{full}
\eea  
The value $\xi=0$ ($\bar \xi= -{1\over 4}$) 
describes the result for a scalar with a minimal coupling. 
A conformally coupled scalar needs instead the 
value $\xi = {(D-2)\over 4(D-1)}$ (i.e. $\bar \xi = {1\over 4(1-D)}$)
together with $m^2 =0$. 
Finally, the value $\xi={1\over 4}$ ($\bar \xi= 0$) allows for  
the simplest computation in the worldline formalism
as all possible vertices contain one graviton only:
this is going to be quite useful for the worldline description of 
spin $1/2$ fermions.

To dispel any further doubt we have repeated the above 
calculations using the standard Feynman rules obtained from
the action in eq. (\ref{uno}) and found the expected 
agreement\footnote{It is likely that the explicit result 
for the two-point function due to a scalar loop is 
present somewhere in the literature, however we have not found it.}.


\section{On the factorization of zero modes}

One can repeat the previous worldline computation using instead the 
propagators with Dirichlet boundary conditions for both coordinates 
and ghosts.
The propagators are different and, in particular, are not translationally 
invariant. 
As a consequence the integrals arising from Wick contractions are more 
complicated and rather laborious to evaluate.
These propagators are again given by 
\bea
\la y^\mu (\tau) y^\nu(\sigma)\ra \!\! &=& \!\! 
- 2 T\ \delta^{\mu\nu}\ 
\Delta(\tau,\sigma)
\nonumber \\ 
\la a^\mu(\tau) a^\nu(\sigma)\ra \!\! &=& \!\! 
 2T\ \delta^{\mu\nu}\ \Delta_{gh}
(\tau,\sigma) \label{propagdbc}
\nonumber \\ 
\la b^\mu(\tau) c^\nu (\sigma)\ra \!\! &=& \!\! 
-4T\ \delta^{\mu\nu}\ \Delta_{gh}(\tau,\sigma)
\eea
but with Green functions $\Delta$ and $\Delta_{gh}$ satisfying 
vanishing DBC 
\bea
\Delta (\tau,\sigma)  \!\! &=& \!\! 
 \sum_{m=1}^{\infty}
\biggl [ - {2 \over {\pi^2 m^2}} \sin (\pi m \tau)
\sin (\pi m \sigma) \biggr ] =
(\tau-1)\sigma\, \theta(\tau-\sigma)+(\sigma-1)\tau\, \theta(\sigma-\tau)
\nonumber \\
\Delta_{gh}(\tau,\sigma) \!\! &=& \!\! 
\sum_{m=1}^{\infty}
2 \, \sin (\pi m \tau) \sin (\pi m \sigma) = \delta(\tau,\sigma)
\label{dbc2}
\eea
where $\theta(\tau-\sigma)$ is the standard step function and 
$\delta(\tau,\sigma)$ is the Dirac's delta function vanishing at the 
boundaries.
Note again that these functions are not translationally invariant. 
Moreover the values of $\Delta$ and $\ddel$  
at coinciding points (which we denote by 
$\Delta|_\tau$ and $\ddel|_\tau$)  
are non-vanishing and, in fact, not even constant. 
One must keep track of them
in the integrals obtained after Wick contractions.
For example, the complete form of the integrals written in 
eq. (\ref{integrals}) for the two-point function in the PBC
method are given in general by
\bea
I_1 \!\! &=& \!\! 
\int_0^1 \!\! d\tau \!\! \int_0^1 \!\! d\sigma\ 
(\ddeld + \Delta_{gh})|_\tau \, (\ddeld + \Delta_{gh})|_\sigma 
\, e^{ T p^2 (\Delta|_\tau + \Delta|_\sigma -2\Delta)}
\nonumber \\ [1mm]
I_2 \!\! &=& \!\! 
\int_0^1 \!\! d\tau \!\! \int_0^1 \!\! d\sigma\ 
(\ddeld{}^2  -\Delta_{gh}^2)
\, e^{ T p^2 (\Delta|_\tau + \Delta|_\sigma -2\Delta)}
\nonumber \\ [1mm]
I_3 \!\! &=& \!\! 
\int_0^1 \!\! d\tau \!\! \int_0^1 \!\! d\sigma\ 
(\ddel - \ddel|_\tau) \, \ddeld \, (\deld  - \deld|_\sigma) 
\, e^{ T p^2 (\Delta|_\tau + \Delta|_\sigma -2\Delta)}
\nonumber \\ [1mm]
I_4 \!\! &=& \!\! 
 \int_0^1 \!\! d\tau \!\! \int_0^1 \!\! d\sigma\ 
(\ddeld + \Delta_{gh})|_\tau\, (\deld -\deld|_\sigma)^2 \, 
e^{ T p^2 (\Delta|_\tau + \Delta|_\sigma -2\Delta)}
\nonumber \\ [1mm]
I_5 \!\! &=& \!\! 
 \int_0^1 \!\! d\tau \!\! \int_0^1 \!\! d\sigma\ 
(\ddel -\ddel|_\tau)^2 \,
(\deld -\deld|_\sigma)^2 \,
e^{ T p^2 (\Delta|_\tau + \Delta|_\sigma -2\Delta)} \ .
\eea
Note that the expressions in (\ref{integrals}) 
are immediately recovered when using the properties of the PBC propagators.
Dimensional regularization can still be used to compute these integrals,
but their values with the DBC propagators differs from the ones 
reported in (\ref{values}). This is correct, as 
there are additional terms that must be included.
In fact, the ghosts have  
vanishing boundary conditions at $\tau=0,1$ and so they cannot create
the covariant measure $\sqrt{ g(x_0)}$ for the integration over the
base-point $x_0^\mu$.
This factor must appear directly in the path integral measure 
\bea
 Dx= {1\over (4 \pi T)^{D \over 2}}\, 
d^Dx_0\,\sqrt{ g(x_0)}\, Dy 
\label{mesplit2}
\eea
and generates additional terms (compare with eq. (\ref{mesplit})).
Suffice here to mention that the extra contributions coming from
the factor $\sqrt{ g(x_0)}$
(there are quadratic terms in $h_{\mu\nu}$
from the direct expansion of $\sqrt{ g(x_0)}$
and a cross term from $\sqrt{ g(x_0)}$ and the action) are
essential to recover the final answer reported in eq. (\ref{full}).
It is quite simple to verify this for the tadpole.
The full contribution comes from the expansion of the factor 
$\sqrt{ g(x_0)}$, while the remaining term from the path integral
vanishes since
\bea
\int_0^1 d\tau\, (\ddeld + \Delta_{gh})|_\tau = 
\int_0^1 d\tau\, \partial_\tau (\deld|_\tau) = 0
\eea
as one verifies from eq. (\ref{dbc2}) using dimensional 
regularization\footnote{This identity is valid also in mode 
regularization, but not in time slicing.
It is easy to check that also these other regularizations produce the correct 
value of the tadpole.} 
(compare this result with eq. (\ref{onepoint})). 
As a consequence we conclude that the string inspired approach 
is more efficient for computing correlation functions.

On the other hand, if one is interested in computing directly
the effective action the opposite is true.
In fact, non-covariant total derivative terms which seem to arise 
in the string inspired approach make the choice of 
Riemann normal coordinates not useful at all \cite{Schalm:1998ix}.
To test these expectations we now compute the leading terms of the effective 
action (in the proper time expansion) using both schemes,
and exhibit the total derivative term arising at the leading order.
The computation is structurally the same as the one carried out in the heat
kernel approach by DeWitt \cite{DW}.
The difference is that instead of solving the heat equation 
with an ansatz to find recursive relations for the so-called 
Seeley-DeWitt coefficients, we compute those coefficients directly 
with a worldline path integral.

We start from the effective action written as in eq. (\ref{effaction3}).
Using arbitrary coordinates we get the following expansion
\bea
\Gamma[g] = -{1\over 2} \int_0^\infty {dT\over T } 
{ e^{-m^2 T} \over (4 \pi T)^{D \over 2}}\,
Z(T) 
\eea
where $Z(T)$ is given by
\bea
Z(T) = \int
d^Dx_0\,\sqrt{ g(x_0)}\, \
\Big \la 1- S_3 - S_4 + {1\over 2}S_3^2 +...\Big \ra
\label{eaone}
\eea
with the vertices 
\bea
S_3 \!\! &=& \!\!  {1\over T}
\int_{0}^{1}\!\!\!  d\tau\ \Bigl [
{1\over 4}\partial_\alpha  g_{\mu\nu} (x_0)\,
y^\alpha 
(\dot y^\mu \dot y^\nu + a^\mu a^\nu +b^\mu c^\nu)\Bigr ]  
\nonumber \\ [1mm]
S_4 \!\! &=& \!\!  {1\over T}
\int_{0}^{1}\!\!\!  d\tau\ \Bigl [
{1\over 8}\partial_\alpha  \partial_\beta  g_{\mu\nu} (x_0)\, 
y^\alpha y^\beta 
(\dot y^\mu \dot y^\nu + a^\mu a^\nu +b^\mu c^\nu)
+ T^2 \bar \xi R (x_0) 
\Bigr ]  \ .
\eea
Notice that we expand around the fixed point $x_0$ 
(the ``base-point'' in the DBC method and the ``zero-mode point''
in the PBC method) but keep the metric arbitrary.
Thus all propagators carry the inverse metric evaluated at that point
\bea
\la y^\mu (\tau) y^\nu(\sigma)\ra = - 2 T\ g^{\mu\nu}(x_0)\ 
\Delta(\tau,\sigma)
\eea
and similarly for the ghosts.

We use a short hand notation for the various tensor 
structures appearing after the Wick contractions (see appendix). We find that 
$\langle S_3 \rangle$
vanishes as it contains an odd number of fields and
\bea 
 \langle -  S_4  \rangle \!\! &=& \!\! 
-{T\over 2} \biggl [ A_1 g^{\mu\nu}\partial^2 g_{\mu\nu}  + 
2 A_2\partial^\mu  \partial^\nu g_{\mu\nu}  \biggr ]  -T  \bar \xi R 
\nonumber  \\ [1mm]
\biggl \langle  {1\over 2} S_3^2 \biggr \rangle \!\! &=& \!\! 
- {T\over 4} \biggl [ B_1 (g^{\mu\nu}\partial_\alpha g_{\mu\nu})^2 
+4 B_2 (g^{\mu\nu}\partial^\alpha g_{\mu\nu})(\partial^\beta g_{\beta\alpha})
+ 2 B_3 (\partial_\alpha g_{\mu\nu})^2 
\nonumber \\ [1mm]
\!\! &+& \!\! 
4 B_4 (\partial^\mu g^{\nu\alpha})(\partial_\nu g_{\mu\alpha}) 
+ 4 B_5 (\partial^\mu g_{\mu\nu})^2
\biggr ]
\label{expa}
\eea
with all tensor structures evaluated at $x_0$,
and with
\bea    
A_1 \!\! &=& \!\! 
\int_{0}^{1} \!\!\! d\tau \ 
\del|_\tau  \ (\ddeld + \Delta_{gh})|_\tau 
\nonumber          
\\
A_2 \!\! &=& \!\! 
\int_{0}^{1}  \!\!\! d\tau \
 \ddel^2|_\tau 
\nonumber        
\\            
B_1 \!\! &=& \!\! 
\int_{0}^{1}  \!\!\!  d\tau \! \int_{0}^{1}  \!\!\!  d\sigma \ 
(\ddeld + \Delta_{gh})|_\tau \ \del \ (\ddeld + \Delta_{gh})|_\sigma
\nonumber
\\               
B_2 \!\! &=& \!\! 
\int_{0}^{1}  \!\!\!  d\tau \! \int_{0}^{1}  \!\!\!  d\sigma \ 
(\ddeld + \Delta_{gh})|_\tau \ \deld \ \deld|_\sigma 
\nonumber
\\               
B_3 \!\! &=& \!\! 
\int_{0}^{1}  \!\!\!  d\tau \! \int_{0}^{1}  \!\!\!  d\sigma \ 
\del \ (\ddeld{}^2 - \Delta_{gh}^2)
\nonumber
\\               
B_4 \!\! &=& \!\! 
\int_{0}^{1}  \!\!\!  d\tau \! \int_{0}^{1}  \!\!\!  d\sigma \ 
 \deld \ \ \ddel \ \ \ddeld 
\nonumber
\\               
B_5 \!\! &=& \!\! 
\int_{0}^{1}  \!\!\!  d\tau \! \int_{0}^{1}  \!\!\!  d\sigma \ 
\deld |_\tau \ \ddeld \ \deld|_\sigma \ .
\eea               

Using the DBC propagators in eq. (\ref{dbc2}) together with
dimensional regularization gives the following values
\bea
A_1= -{1\over 6}  \ , \ \
A_2= {1\over 12}  \ , \ \
B_1= -{1\over 12}  \ , \ \
B_2= {1\over 12}  \ , \ \
B_3= {1\over 8}  \ , \ \
B_4= -{1\over 24}  \ , \ \
B_5= -{1\over 12}  
\eea
so that with the help of formula (\ref{help}) in the appendix
one can cast the final result as
\bea
Z_{DBC}(T) = 
\int d^Dx\,\sqrt{ g} \Big [ 1 - T 
\Big ({ 1\over 12 }+\bar \xi \Big ) R
+ O(T^2) \Big ]
\ .
\eea

On the other hand, using the PBC propagators in eq. (\ref{ghost-prop}) 
gives 
\bea
A_1= -{1\over 12}  \ , \ \
A_2= 0  \ , \ \
B_1= 0 \ , \ \
B_2= 0 \ , \ \
B_3 = {1\over 24} \ , \ \  
B_4= {1\over 24}  \ , \ \
B_5= 0
\eea
and inserting these values into (\ref{expa}) produces 
\bea
Z_{PBC}(T) = 
\int d^Dx\ \Big [ \sqrt{ g} 
- T 
\sqrt{ g}
\Big  ({ 1\over 12 }+\bar\xi\Big ) 
R + {T\over 12} \partial_\mu ( \sqrt{g} g^{\alpha\beta}
\Gamma_{\alpha\beta}{}^\mu ) \Big ] + O(T^2)\ .
\eea
We see explicitly the total derivative term appearing at this 
perturbative order in the PBC method. It is manifestly 
non-covariant and reads
\bea
\Delta Z(T) = {T\over 12}  \int d^Dx\ 
 \partial_\mu ( \sqrt{g} g^{\alpha\beta}
\Gamma_{\alpha\beta}{}^\mu ) + O(T^2) \ .
\eea
It implies that Riemann normal coordinates 
would not be useful to simplify the calculations in 
the PBC (``string inspired'') method: 
one would not know how to reconstruct the final
expression in arbitrary coordinates.
We interpret this total derivative term as an infrared effect due to the 
non-local constraint
$\int_0^1 d\tau\, y^\mu(\tau)=0$ imposed on the quantum fields. 
Of course, the same total derivative term
is found using mode regularization
or time slicing\footnote{In fact one of us (FB) 
originally computed this term in a discussion with Christian Schubert
using only MR and TS regularizations.}.

\section{Conclusions}

We have discussed the worldline formalism for a scalar particle
coupled to a gravitational background.
Using an ultraviolet regularization we have shown how the
results expected from QFT follow unambiguously.
We have seen that the easiest way to proceed for calculating
correlation functions at one-loop 
is to use: {\it (i)} 
periodic boundary conditions with factorization of the zero
mode and {\it (ii)} 
worldline dimensional regularization.
With these prescriptions we have computed the scalar particle contribution
to the graviton tadpole and self-energy.
We have also seen explicitly that the PBC method suffers from non-covariant
total derivative terms arising when one wants 
to compute the effective action directly. 
One may hope that future improvements will show a way
of using the PBC method for effective action calculations, 
but at present the secure path is to use the DBC method if one wants 
to employ the simplifying properties of Riemann normal coordinates.

We expect that the worldline formalism with gravity can be extended 
to include other types of particles in the loop. 
More ambitiously one would like to
establish a connection with the string inspired rules of 
refs. \cite{Bern:1993wt,Bern:1998ug}.
\vskip 1cm 

{\bf Acknowledgments}

We thank Christian Schubert for discussions and comments.

\vfill \eject

\begin{appendix} 
\section*{Appendix}
We use the following conventions for the curvature tensors
\bea
[\nabla_\mu, \nabla_\nu] V^\lambda = 
R_{\mu\nu}{}^\lambda{}_\rho V^\rho \ , \ \ \ 
R_{\mu\nu}= R_{\lambda\mu}{}^\lambda{}_\nu 
\ , \ \ \ R= R^\mu{}_\mu > 0 \ {\rm on\ spheres.} 
\eea
In a self-evident, short-hand notation the scalar curvature is given by 
the sum of the following 7 terms
\bea
R \!\! &=& \!\! 
\partial^\mu \partial^\nu g_{\mu\nu}-g^{\mu\nu}\partial^2 g_{\mu\nu} 
+{3\over 4} (\partial_\alpha g_{\mu\nu} )^2 
-{1\over 2}(\partial^\mu g^{\nu\alpha} )(\partial_\nu g_{\mu\alpha})
-{1\over 4} (g^{\mu\nu} \partial_\alpha g_{\mu\nu} )^2
\nonumber \\ 
\!\! &+& \!\! 
 (g^{\mu\nu} \partial^\alpha g_{\mu\nu}) (\partial^\beta g_{\beta\alpha})
- (\partial^\mu g_{\mu\nu})^2 \ . 
\label{help}
\eea
The linear and quadratic terms in the expansion 
around flat space read as 
\bea
R \!\! &=& \!\!  R^{(1)} + R^{(2)} + ...\nonumber \\
R^{(1)} \!\! &=& \!\!  \partial^\mu \partial^\nu h_{\mu\nu} - \square h
\nonumber \\  
R^{(2)} \!\! &=& \!\!  h^{\mu\nu}\square  h_{\mu\nu} 
- 2 h^{\mu\nu}\partial_\mu 
\Big (\partial^\alpha h_{\alpha\nu} - {1\over 2} \partial_\nu h \Big) 
- \Big(\partial^\alpha h_{\alpha\mu} - {1\over 2} \partial_\mu h \Big)^2
+{3\over 4} (\partial_\alpha h_{\mu\nu})^2 \nonumber \\
&-& \!\! {1\over 2} (\partial_\mu h_{\nu\alpha})(\partial^\nu h^{\mu\alpha}) 
\label{r2}
\eea 
where $h_{\mu\nu}=g_{\mu\nu}-\delta_{\mu\nu}$
and $h =\delta^{\mu\nu}h_{\mu\nu}$.
\end{appendix} 

\vfill \eject



\begin{thebibliography}{99}
\baselineskip=14pt
 
\bibitem{Feynman50}
R.~P.~Feynman,
Phys.\ Rev.\  {\bf 80} (1950) 440;
J.~S.~Schwinger,
Phys.\ Rev.\  {\bf 82} (1951) 664;
Y.~Nambu,
Prog.\ Theor.\ Phys.\  {\bf 5}, 82 (1950);
V.~Fock,
Phys.\ Z.\ Sowjetunion {\bf 12} (1937) 404.

\bibitem{BK}
Z.~Bern and D.~A.~Kosower,
Phys.\ Rev.\ Lett.\  {\bf 66} (1991) 1669;
Nucl.\ Phys.\ B {\bf 379} (1992) 451;
Z.~Bern, L.~J.~Dixon and D.~A.~Kosower,
Phys.\ Rev.\ Lett.\  {\bf 70} (1993) 2677
[hep-ph/9302280].


\bibitem{Polyakov} 
A.M. Polyakov, {\em ``Gauge Fields and Strings''} 
(Harwood, Chur, Switzerland, 1987).


\bibitem{Strassler}
M.J. Strassler, Nucl.\ Phys.\ B {\bf 385} (1992) 145 [hep-th/9205205];\\
M.~G.~Schmidt and C.~Schubert,
Phys.\ Lett.\ B {\bf 318} (1993) 438
[hep-th/9309055];
Phys.\ Lett.\ B {\bf 331} (1994) 69
[hep-th/9403158];\\
E.~D'Hoker and D.~G.~Gagne,
Nucl.\ Phys.\ B {\bf 467} (1996) 272
[hep-th/9508131];
Nucl.\ Phys.\ B {\bf 467} (1996) 297
[hep-th/9512080].


\bibitem{Chris}
C.~Schubert,
Phys.\ Rept.\  {\bf 355} (2001) 73
[hep-th/0101036].


\bibitem{Bern:1993wt}
Z.~Bern, D.~C.~Dunbar and T.~Shimada,
Phys.\ Lett.\ B {\bf 312} (1993) 277
[hep-th/9307001];
D.~C.~Dunbar and P.~S.~Norridge,
Nucl.\ Phys.\ B {\bf 433} (1995) 181
[hep-th/9408014].

\bibitem{Brink:1976sz}
L.~Brink, S.~Deser, B.~Zumino, P.~Di Vecchia and P.~Howe,
Phys.\ Lett.\ B {\bf 64} (1976) 435.


\bibitem{Bastianelli:1992be}
F.~Bastianelli,
Nucl.\ Phys.\  B {\bf 376} (1992) 113
[hep-th/9112035].

\bibitem{Bastianelli:1993ct}
F.~Bastianelli and P.~van Nieuwenhuizen,
Nucl.\ Phys.\  B {\bf 389} (1993) 53
[hep-th/9208059].

\bibitem{Bastianelli:2000dw}
F.~Bastianelli and O.~Corradini,
Phys.\ Rev.\ D {\bf 63} (2001) 065005
[hep-th/0010118];
F.~Bastianelli and N.~D.~Hari Dass,
Phys.\ Rev.\ D {\bf 64} (2001) 047701
[hep-th/0104234].

\bibitem{Bastianelli:1998jm}
F.~Bastianelli, K.~Schalm and P.~van Nieuwenhuizen,
Phys.\ Rev.\ D {\bf 58} (1998) 044002
[hep-th/9801105];
F.~Bastianelli and O.~Corradini,
Phys.\ Rev.\  D {\bf 60} (1999) 044014
[hep-th/9810119].

\bibitem{deBoer:1995hv}
J.~de Boer, B.~Peeters, K.~Skenderis and P.~van Nieuwenhuizen,
Nucl.\ Phys.\ B {\bf 446} (1995) 211 
[hep-th/9504097];
Nucl.\ Phys.\ B {\bf 459} (1996) 631 
[hep-th/9509158].


\bibitem{Bastianelli:2000nm}
F.~Bastianelli, O.~Corradini and P.~van Nieuwenhuizen,
Phys.\ Lett.\ B {\bf 494} (2000) 161
[hep-th/0008045].

\bibitem{KC}
H. Kleinert and A. Chervyakov,
Phys. Lett. B {\bf 464} (1999) 257
[hep-th/9906156];
F. Bastianelli, O. Corradini and P. van Nieuwenhuizen,
Phys. Lett. B {\bf 490} (2000) 154
[hep-th/0007105].

\bibitem{Schalm:1998ix}
K.~Schalm and P.~van Nieuwenhuizen,
Phys.\ Lett.\ B {\bf 446} (1999) 247
[hep-th/9810115].

\bibitem{FT}
E.~S.~Fradkin and A.~A.~Tseytlin,
Phys.\ Lett.\ B {\bf 158} (1985) 316;
Nucl.\ Phys.\ B {\bf 261} (1985) 1.


\bibitem{Fliegner:1997rk}
D.~Fliegner, P.~Haberl, M.~G.~Schmidt and C.~Schubert,
Annals Phys.\  {\bf 264} (1998) 51
[hep-th/9707189].


\bibitem{Bern:1998ug}
Z.~Bern, L.~J.~Dixon, D.~C.~Dunbar, M.~Perelstein and J.~S.~Rozowsky,
Nucl.\ Phys.\ B {\bf 530} (1998) 401
[hep-th/9802162],
Z.~Bern,
in {\it Proc. of the 5th International Symposium on Radiative Corrections 
(RADCOR 2000) } ed. Howard E. Haber [hep-th/0102186].

\bibitem{ly} 
T.~D.~Lee and C.~N.~Yang,
Phys.\ Rev.\  {\bf 128} (1962) 885;
E.~S.~Abers and B.~W.~Lee,
Phys.\ Rept.\  {\bf 9} (1973) 1.


\bibitem{DW} B.S. DeWitt, in ``{\em Relativity, Groups and Topology}''
(lectures at Les Houches 1963) ed. B. and C. DeWitt (Gordon Breach, NY, 1964);
``{\em Relativity, Groups and Topology II}'' (lectures at Les Houches 1983)
ed. B. DeWitt and R. Stora (North Holland, Amsterdam, 1984).

\end{thebibliography}
\end{document}